\documentclass[conference]{IEEEtran}
\IEEEoverridecommandlockouts

\pdfoutput=1

\usepackage{graphicx}
	\graphicspath{{Figures/}}
	\DeclareGraphicsExtensions{.pdf}
	
\usepackage[cmex10]{amsmath}
\usepackage{amsfonts}
\usepackage{xcolor,colortbl}
\usepackage[]{algorithm2e}
\usepackage{stfloats}
\usepackage{siunitx}

\begin{document}

\title{Geometry-Based Stochastic Channel Models for 5G: Extending Key Features for Massive MIMO}
\author{\IEEEauthorblockN{{\`A}lex Oliveras Mart{\'i}nez, Patrick Eggers, Elisabeth De Carvalho}
\IEEEauthorblockA{Faculty of Engineering and Science, Dept. of Electronic Systems, APNet section\\
Aalborg University,
Aalborg, Denmark\\ 
Email: \{aom,pe,edc\}@es.aau.dk}}

\maketitle

\begin{abstract}
This paper introduces three key features in geometry-based
stochastic channel models in order to include massive MIMO channels.
Those key features consists of multi-user (MU) consistency, non-stationarities across the base station array and inclusion of spherical wave modelling. 
To ensure MU consistency, we introduce the concept of ``user aura'', which is a circle around the user with radius 
 defined according to the stationarity interval. 
The overlap between auras determines the share of common clusters among users. 
To model non-stationarities across a massive array, sub-arrays are defined for which clusters are independently generated. At last, we describe a procedure to incorporate spherical wave modelling, where a cluster focal point is defined to account for distance between user and cluster.

\end{abstract}

\IEEEpeerreviewmaketitle

\section{Introduction}
{I}n a {massive} MIMO (Multiple-Input Multiple-Output) system, the base station is equipped with a very large number of antenna elements and serves multiple users in the same time-frequency resource~\cite{Marzetta2010Noncooperative}. Under certain favorable propagation conditions (e.g.~\cite{Ngo2014Aspects}), fast fading  and uncorrelated noise at the receiver  vanish, bringing huge gains in throughput, reliability and energy efficiency~\cite{Larsson2014Massive}. Massive MIMO is considered a key technology for the development of 5G~\cite{Boccardi2014Five}.

The characteristics of the massive MIMO channel bring some challenges for inclusion in the existing geometry-based stochastic channel models (GSCM). The existing GSCM can be divided into two groups. We name them Winner-type and COST-type. The first ones are the main focus of this work and examples are the 3GPP spatial channel model (SCM), extended SCM (SCME)~\cite{Baum2005}, Winner (WIM1), Winner II (WIM2)~\cite{Svensson2007}, Winner+ (WIM+) and QuaDRiGa~\cite{Jaeckel2014}. Their main characteristic consists of the definition of the scatterers based on the angles of departure and angles of arrival, i.e. terminal perspective. On the other hand COST-type GSCM~\cite{Liu2012} defines the physical position of the scatterers in the simulation area.

Existing work proposes an extension of COST-type GSCM for massive MIMO~\cite{Gao2015Extension}. The COST-type GSCM channel models defines the physical position of the scatterers, not directly angles of departure or arrival as seen from terminal. Consequently, it is difficult to extract parameters for the COST model using measurements (contrary to the case of the Winner type GSCM channel model). Those reasons explain why Winner-type GSCM  is currently more widespread and is the preferred candidate for 5G channel modelling in standardization efforts. 


One major drawback of Winner-type GSCM is that it does not support multi-user (MU) consistency. 
MU consistency refers to the generation of channels for each users that are consistent with the distance between users in terms of observed clusters and their correlation.
Winner-type GSCM  fails to represent scenarios where the users are in close proximity, as the channels are generated independently for each user,  regardless of the distance between users. 
 As the performance of massive MIMO is  related to the user channel vectors orthogonality~\cite{Ngo2014Aspects}, Winner-type GSCM models results in over-optimistic performance.

\begin{figure}[!t]
\centering
\fbox{\includegraphics[trim = 2.3in 2.7in 2.3in 2.6in, clip, width=3.3in]{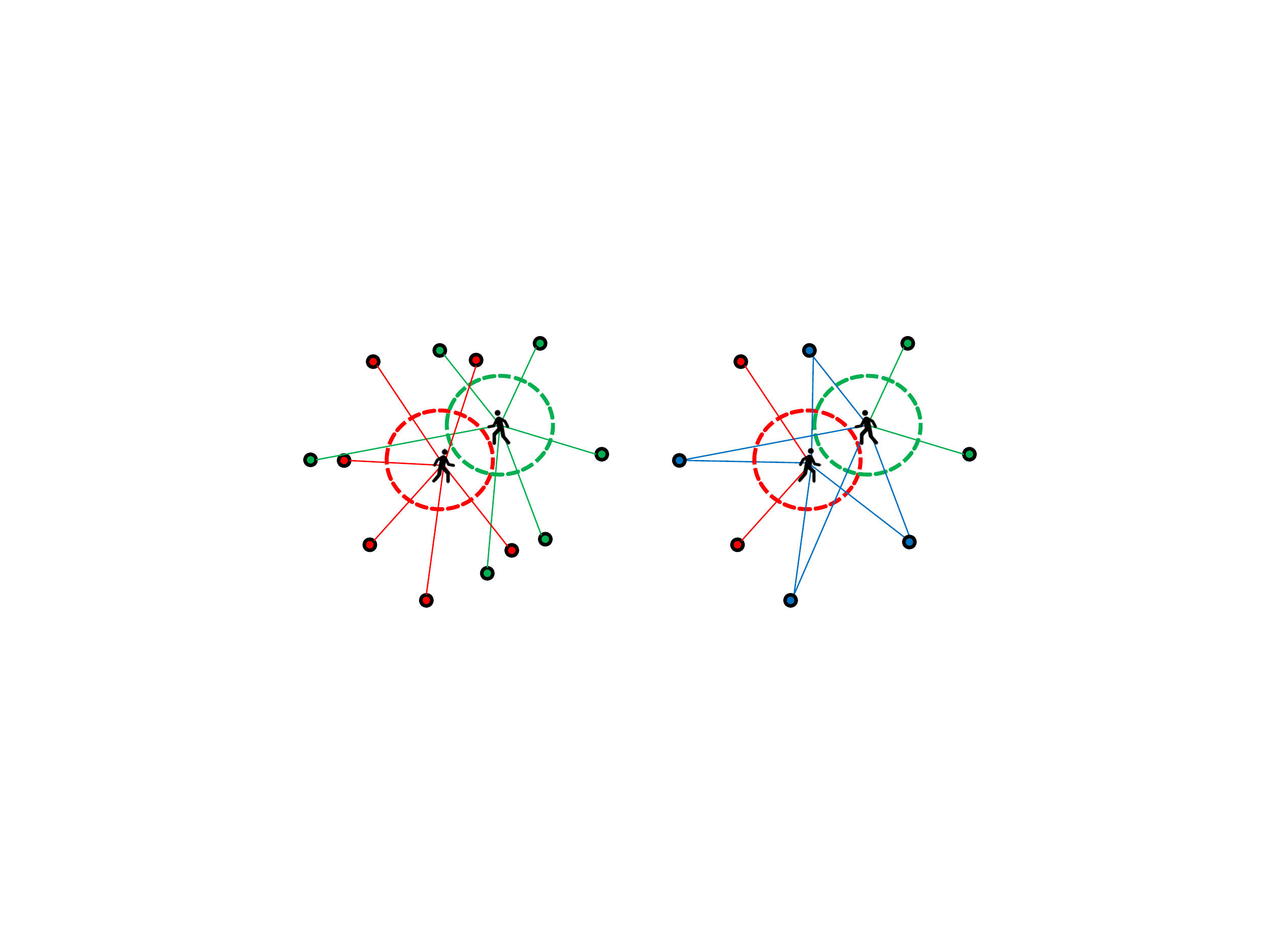}}
\caption{Left: Existing GSCM with independent clusters for users closely located. Right: Proposed extension with multiuser consistency where closely located users have common clusters}
\label{fig_User_Consistency}
\end{figure}

With the increase in the number of antennas, the size of the arrays also increases. Although compact array designs are desirable for operators, some papers argue that the real advantages of massive MIMO appear when the size of the array become large~\cite{Martinez2014Towards}. Non-stationarities have been observed in measurements~\cite{Martinez2014Towards, Gao2012} for large but also compact arrays~\cite{Martinez2014Towards}, so that it would appear that even for compact arrays, it becomes important to model non-stationarities. The non-stationarities are of different nature: the power can vary, the directions of departure/arrival varies, different parts of the array see different clusters, etc.\\ \indent
Winner-type GSCM  define the clusters by their angles of departure and angles of arrival and rely on a planar wave approximation. When the array becomes larger or the clusters are at close proximity to the users, the planar wave approximation becomes inexact calling for a spherical wave modelling and a modification of Winner-type GSCM  models.\\ \indent
The present study proposes solutions to extend Winner-type GSCM to include MU consistency, non-stationarities across the base station array and spherical wave modelling. 
We introduce the concept of ``aura'' associated to each user, which is a circle centered on the user with radius defined by the stationarity interval. When users are at close proximity, their auras overlap and the overlapping surface defines the common clusters shared among users. An example is presented in Fig.~\ref{fig_User_Consistency}. We propose an algorithm to compute the number of clusters to be shared among pairs of users, then groups of multiple users.
As a pre-step,  an algorithm to divide the users into connected groups is used to increase the speed of the process. 
The large scale parameters (LSP) of one of the users sharing the cluster are used to compute the parameters of the cluster.
Finally the cluster parameters are shared with the other users sharing the cluster. Then the parameters or the position of the clusters (depending on the distance between user and cluster) is recalculated according to the position of the new user.

To account for non-stationarity effects, the base station is divided into sub-arrays with size defined by the stationarity distance (i.e. correlation or coherence distance with regard to visible clusters). Different realizations of the LSP at each sub-array are used to generate the clusters. 

Spherical wave modeling, 
(similarly to QuaDRiGa's drifting procedure~\cite{Jaeckel2014}) supports near field clusters at the base station side, by fixing focal points derived from the delays and angles of the clusters.

This work uses QuaDRiGa, a Winner-type GSCM, 
 as a reference model on top of which to build the proposed extensions. However the ideas in the paper can also be applied to other Winner-type GSCM.

\section{Definition of basic concepts}

\subsection{Segments}
From WIM1 onwards the trajectory of the users is divided into smaller segments. These segments are defined such as the LSP of the channel remain constant (i.e. the segment length equals the stationarity interval). Therefore the number of clusters that each user has and the correspondent parameters can only change segment wise. This paper proposes an algorithm that checks the relative position of the users at the beginning of each segment and defines the number of common cluster for the rest of the segment. The proposed extension of the GSCM can only be applied if the user segment transitions are synchronized.

\subsection{Clusters}
\begin{figure}[!t]
\centering
\fbox{\includegraphics[trim = 1.6in 2.2in 1.6in 1.8in, clip, width=3.3in]{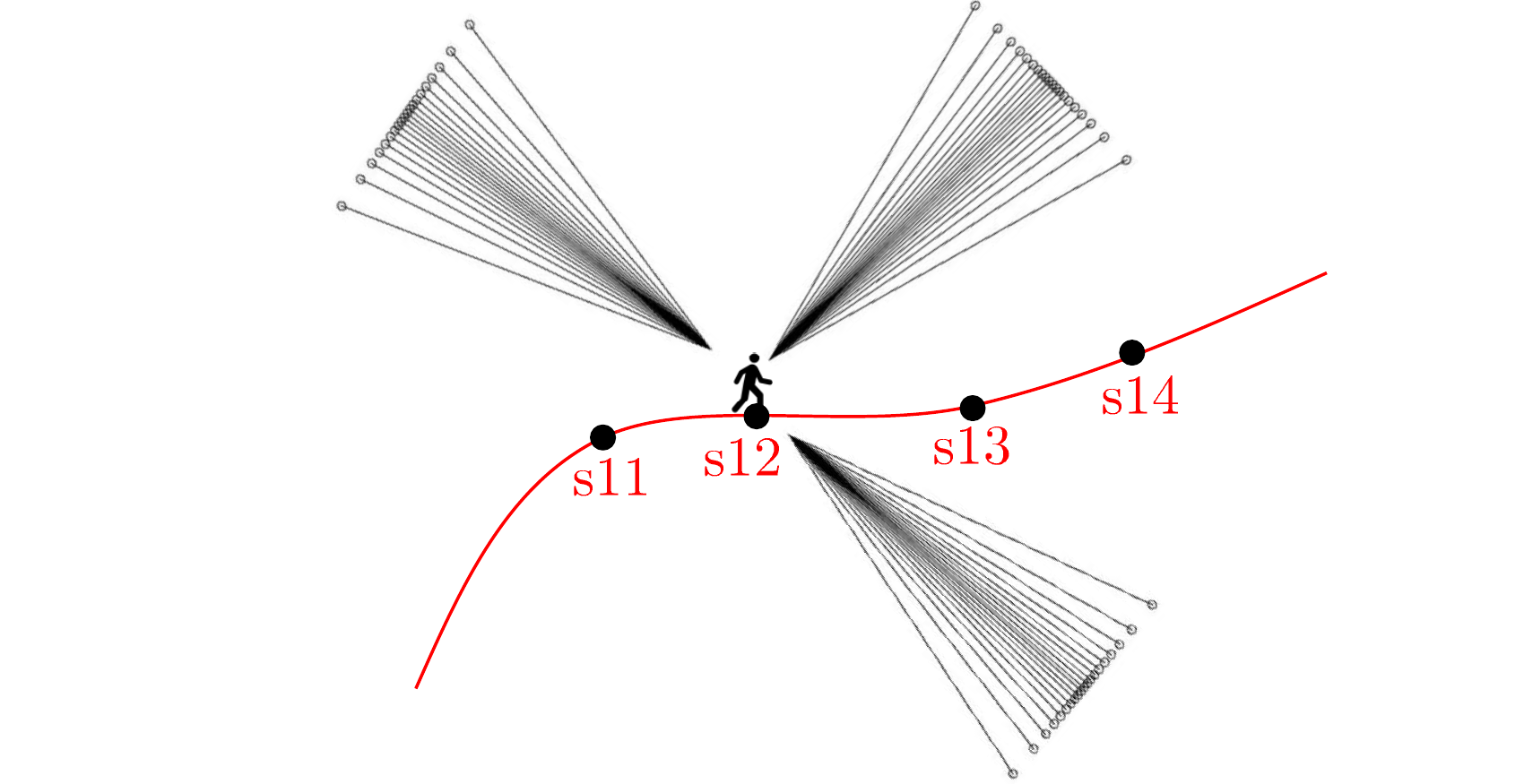}}
\caption{Angular representation of the clusters in Winner-type GSCM. The AoA defines the mean angle of the cluster and the angles of the scatters have a deterministic offset from this angle. All the scatterers in the cluster have the same delay (Not represented in this figure)}
\label{fig_Winner_Cluster_User}
\end{figure}

In GSCM, physical objects are modeled as scatterers where the transmitted waves are reflected. These scatterers are divided into groups according to their delay and angle of departure or arrival, forming clusters. Each cluster is composed by $20$ scatterers. The angles of the scatterers are samples of a Laplacian function, as shown in Fig.~\ref{fig_Winner_Cluster_User}. The parameters of the clusters of each user define the channel properties of that user. The proposed extension proposes to share some of the clusters between users that are near to each other to achieve the desired multiuser consistency. The number of shared clusters is related to the distance between users, but which clusters are shared can vary according to the implementation.

\begin{figure}[!t]
\centering
\includegraphics[width=3.1in]{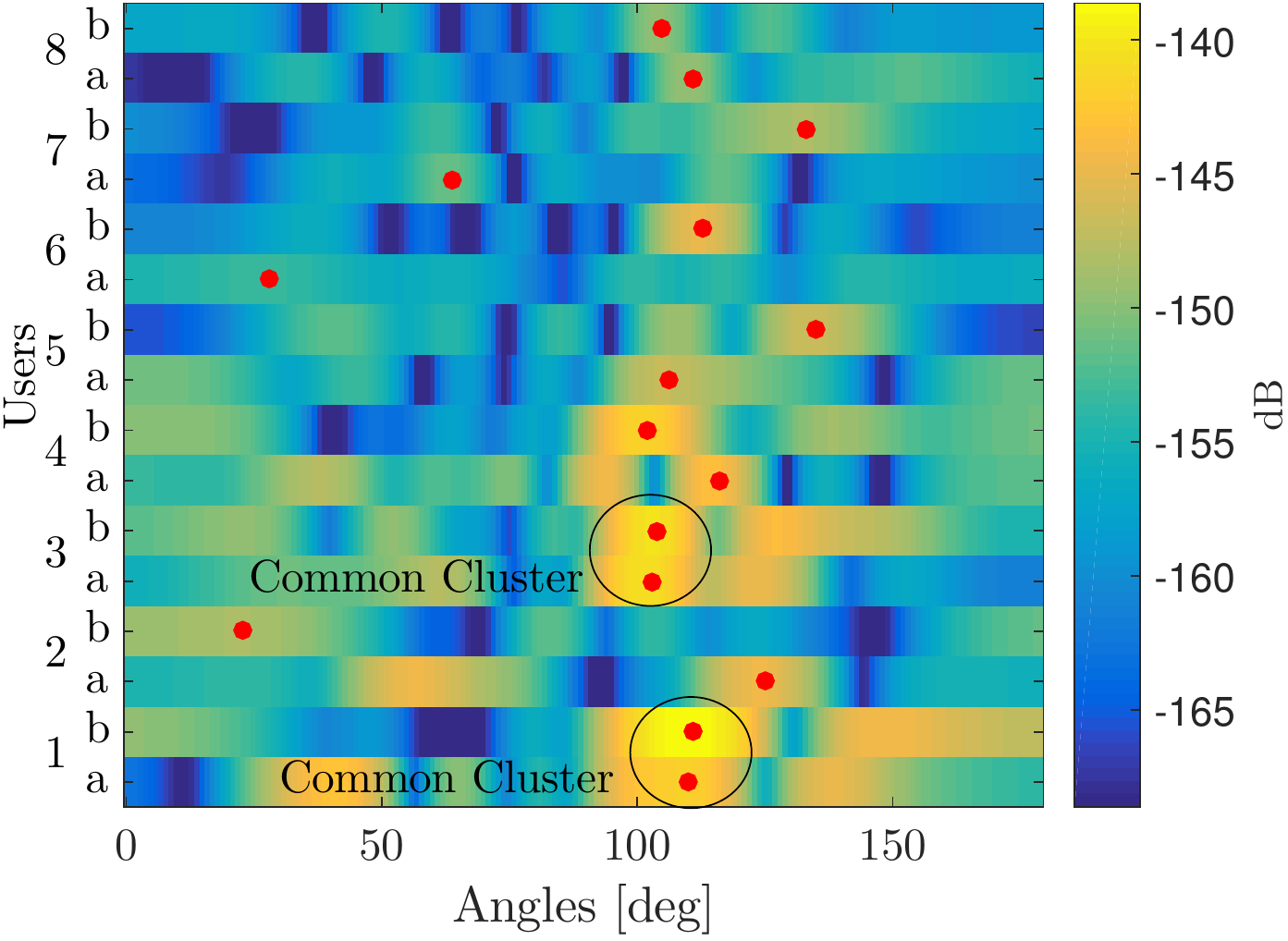}
\caption{Common cluster observed between user $1$ and $3$ in an indoor NLoS scenario. Red dots mark the angle of maximum power}
\label{fig_Common_Cluster}
\end{figure}

The common clusters between users have been observed in a massive MIMO channel measurement described in~\cite{Martinez2014Towards}. Due to lack of space the measurement campaign is not described here (for detail see~\cite{Martinez2014Towards}). In the measurement campaign the base station array has $64$ elements divided in sets of $8$ elements. There are $8$ users holding a handset with $2$ elements called $a$ and $b$. The angle of arrival is estimated using steering-vector beamforming at each set of $8$ elements. The $8$ element set has a \SI{13}{\degree} \SI{-3}{\decibel} beamwidth and max. sidelobe level of \SI{-14.7}{\decibel}. We focus on the maximum power cluster to avoid a misinterpretation of the side lobes. This cluster is marked with a red dot. In Fig.~\ref{fig_Common_Cluster} the power angular spectrum for the $8$ users in a non-Line-of-Sight scenario (called S-NLoS in~\cite{Martinez2014Towards}) is presented. Fig.~\ref{fig_Common_Cluster} shows that user $1$ and user $3$ separated \SI{2.2}{\meter} have a cluster at \SI{104}{\degree} and \SI{110}{\degree} respectively. Due to the \SI{13}{\degree} resolution of the beamforming this can be considered a common cluster.

\subsection{User aura}
\begin{figure}[!t]
\centering
\fbox{\includegraphics[trim = 2.1in 2.8in 2.1in 2.8in, clip, width=3.3in]{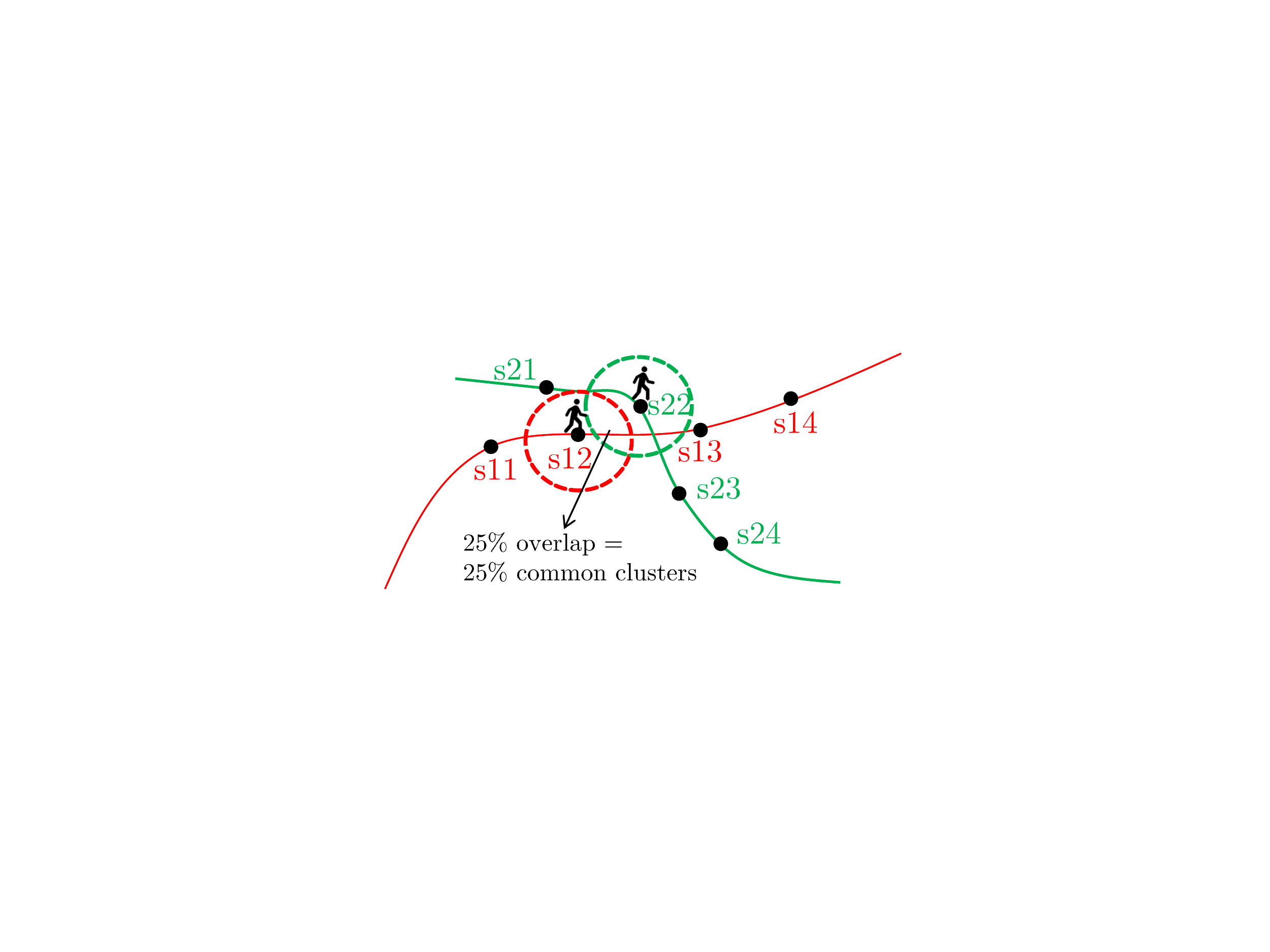}}
\caption{Track of two users divided into segments. sUX is the first position of user U in segment number X}
\label{fig_Overlapping}
\end{figure}

Where COST-type GSCM provide natural cluster sharing as it is cluster centric, the Winner-type GSCM are user centric. Despite QuaDRiGa made a mapping of the parameters to geometric positions to provide time evolution of the channel, it is still local to each user. To facilitate a possible sharing of clusters between users, we introduce the concept of an aura. The user aura is defined as the circle surrounding the users with radius equal to the stationarity interval. When two users are separated more than the stationarity interval they have independent channel vectors and their auras are disjoint. If two users are close to each other their auras overlap. The amount of overlapping area is proportional to the distance between the users. This proportion is used to define the amount of clusters that need to be shared between users. The overlapping of the auras is computed in the first position of each segment and the number of common clusters is kept constant along the segment. Fig.~\ref{fig_Overlapping} shows an example users' layout.

\subsection{Aura at the base station}
\begin{figure}[!t]
\centering
\fbox{\includegraphics[trim = 2.9in 3in 1.8in 2.7in, clip, width=3.3in]{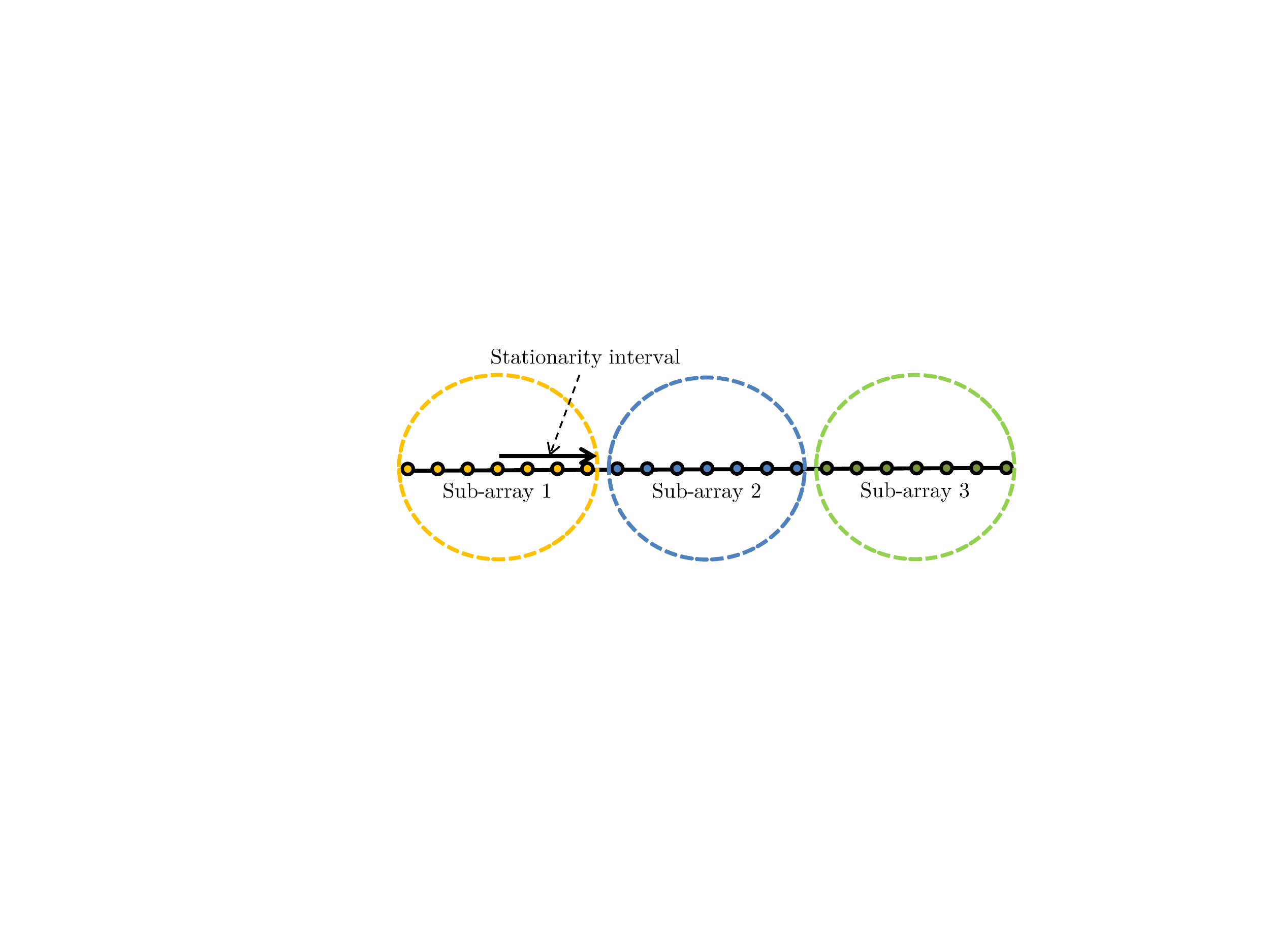}}
\caption{Division of the base station array into sub-arrays according to the stationarity interval}
\label{fig_Sub_array}
\end{figure}

To generate non-stationaries from the base station perspective, the array is divided into sub-arrays in the same way the user trajectory is divided into segments. In the same way each user has defined an aura, the sub-arrays in the base station have also defined an aura of radius equal to the stationarity interval. The sub-arrays have the length of the stationarity interval, and the aura is centered at the center of the sub-array. Notice that the stationarity interval along the base station array might differ from the user perspective. To generate the non-stationarities over the array, the exposed extension proposes to chose different parameters of the cluster at the transmitter for each sub-array. See an example in Fig.~\ref{fig_Sub_array}. The auras of adjacent sub-arrays can be overlapped to produce a gradual share of the clusters. However, in this work it is not implemented.

\subsection{Drifting}
One contribution of QuaDRiGa is the time evolution channel consistency. It means that in each snapshot the channel is consistent with the previous and the following snapshot. At the initial position of the segment the parameters of the clusters are calculated. These parameters are updated according to the movement of the user in the subsequent snapshots of the channel. To achieve this time evolution the position of the scatterers has to be defined and kept constant through the segment. Then, the relative position between the user and the scatter can be computed, and the new parameters of the scatter can be updated. The proposed extension uses this concept to define the position of the scatters both at the receive and the transmit side and therefore the spherical waves can be used.

\section{Extension of GSCM}
\subsection{Simulation flow}
Winner-type GSCM channel model follows nine steps to generate the channel coefficients~\cite{Jaeckel2014}:
\begin{enumerate}
\item Define the parameters of the simulation (Positions of the users and base stations, Antenna arrays, Tracks of the users, Segments, Scenarios)
\item	Generation of the correlation maps using the scenarios configuration files
\item	Generation of clusters for each segment
\item	Generation of the scatterers inside the clusters and calculation of the vector for each scatterer and each position of the user
\item	Calculate antenna response for each angle
\item	Calculate the phases using the position of the clusters and the antennas
\item	Sum of the coefficients of the $20$ scatterers. The channel matrix for each cluster is created
\item	Merge the adjacent segments (birth/death process)
\item	Formatting of channel coefficients and delays
\end{enumerate}

To obtain the multi-user consistency, non-stationarities across the array and the spherical waves propagation, Aalborg University (AAU) modifies step $3$ of the nine steps.

\subsection{Description of extensions}
\begin{figure}[!t]
\centering
\includegraphics[trim = 2.4in 2.8in 2.1in 2.7in, clip, width=3.3in]{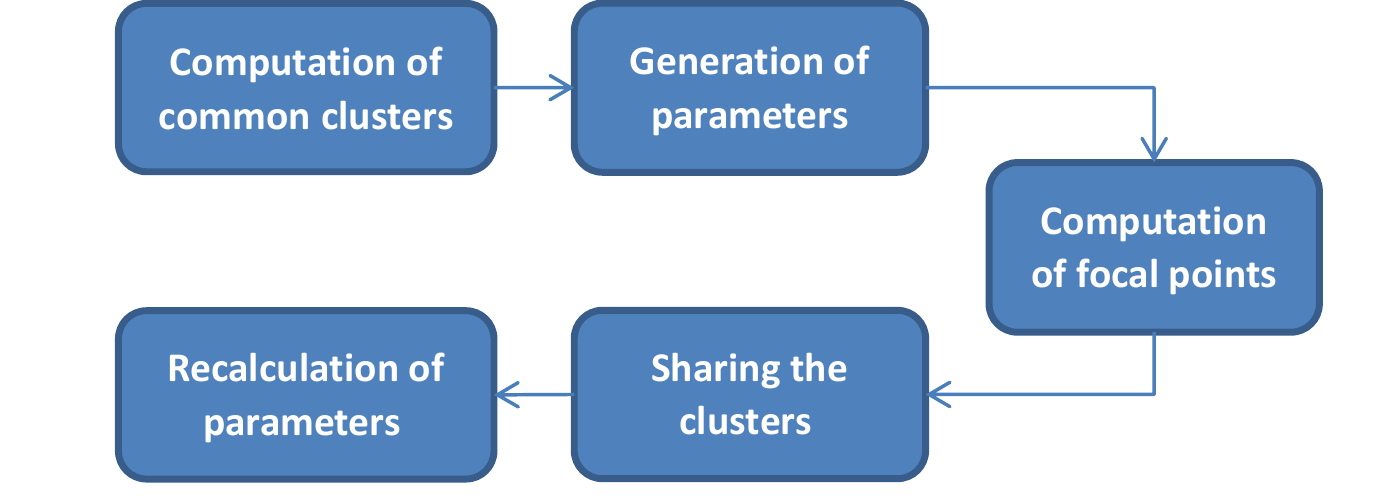}
\caption{Modification of the QuaDRiGa step $3$ into $5$ sub-steps}
\label{fig_Sub_steps}
\end{figure}

The modified step $3$ has $5$ sub-steps. These sub-steps are presented in Fig.~\ref{fig_Sub_steps}:
\begin{enumerate}
\item Calculate proportion of common clusters
\item	Generation of initial parameters
\item	Computation of the focal points of the clusters
\item	Sharing the clusters
\item	Recalculating parameters
\end{enumerate}

\subsection{Calculate proportion of common clusters}
\label{sec_Calculate}
The proposed solution uses a simple preprocessing algorithm and an algorithm designed by AAU to compute the number of common cluster between users depending on their proximity. These algorithms are simple to implement. We want to remark that an algorithm for computing the overlapping of circles with exact precision already exist in~\cite{Librino2009}. The implementation of such algorithm can be complex and the processing time long. The accuracy provided by such algorithm is not necessary. Therefore we develop a simplified method. 

\begin{figure}[!t]
\centering
\includegraphics[trim = 1.6in 2.7in 1.2in 2.3in, clip, width=3.3in]{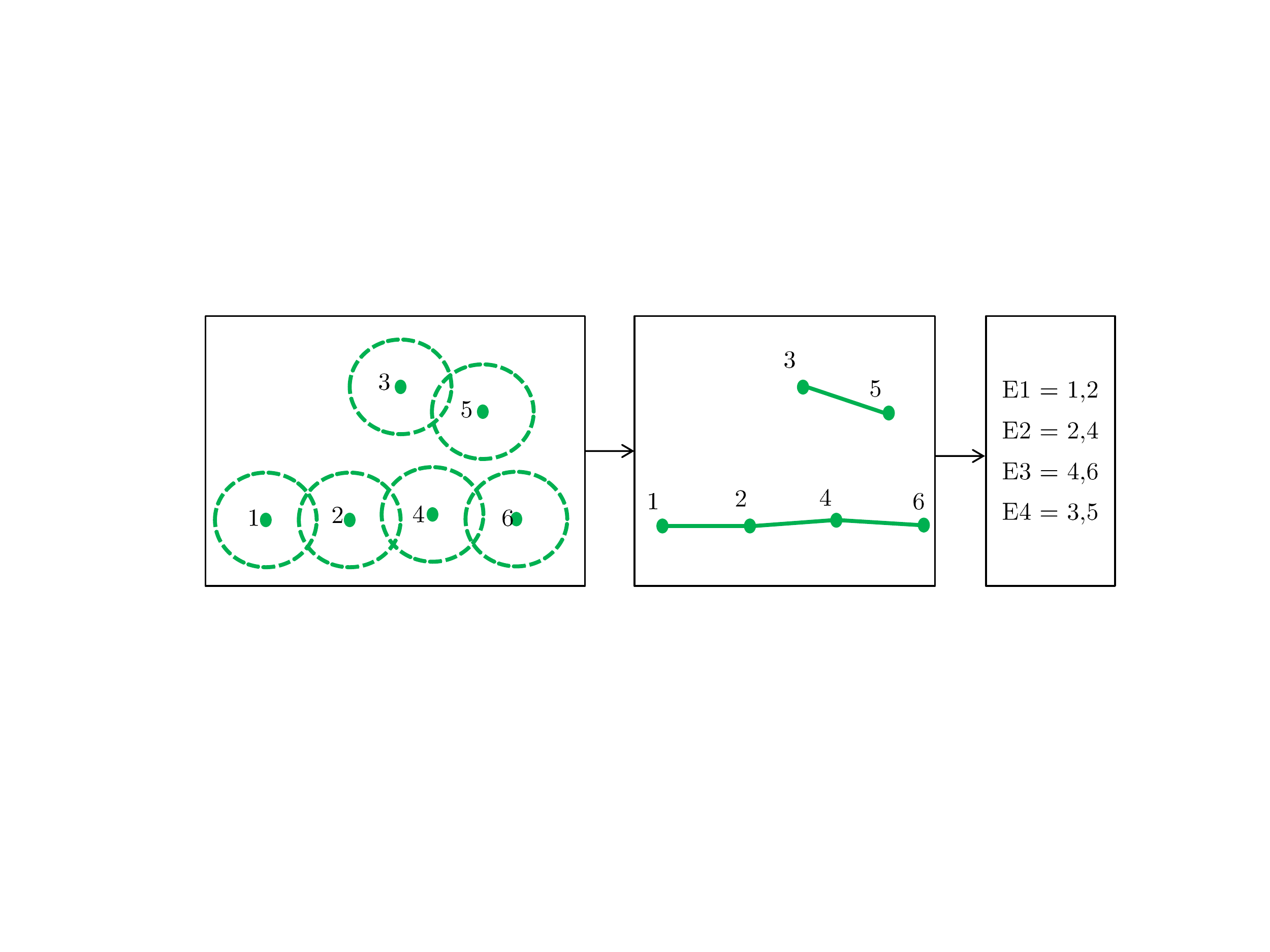}
\caption{Simulation layout with six users converted into a graph and an edges description}
\label{fig_Clustering_Algorithm}
\end{figure}

First there is a preprocessing of the layout to cluster the users in connectivity groups. This algorithm makes groups of users whose auras are overlapping~\cite{Hopcroft1971}. This step is necessary to increase the efficiency of the algorithm to compute the number of common clusters. This algorithm uses graph theory to find connected components. Each user is represented as a vertex of the graph. If the distance between two users is smaller than the sum of its radius their auras are overlapping and there is an edge between the two vertices representing the two users. An example can be seen in Fig.~\ref{fig_Clustering_Algorithm}. The algorithm performs a deep search on each connected component. Each new vertex reached is marked. When no more vertices can be reached along edges from marked vertices, a connected component has been found. An unmarked vertex is then selected, and the process is repeated until the entire graph is explored. This algorithm requires memory space linear with $max(V,E)$, and time linear with $max(V,E)$. Where $V$ is the number of vertices (i.e. users in the layout) and $E$ is the number of edges of the graph (i.e. overlapping auras in the layout).

\begin{figure}[!t]
\centering
\fbox{\includegraphics[trim = 1.5in 2.3in 0.8in 2.3in, clip, width=3.3in]{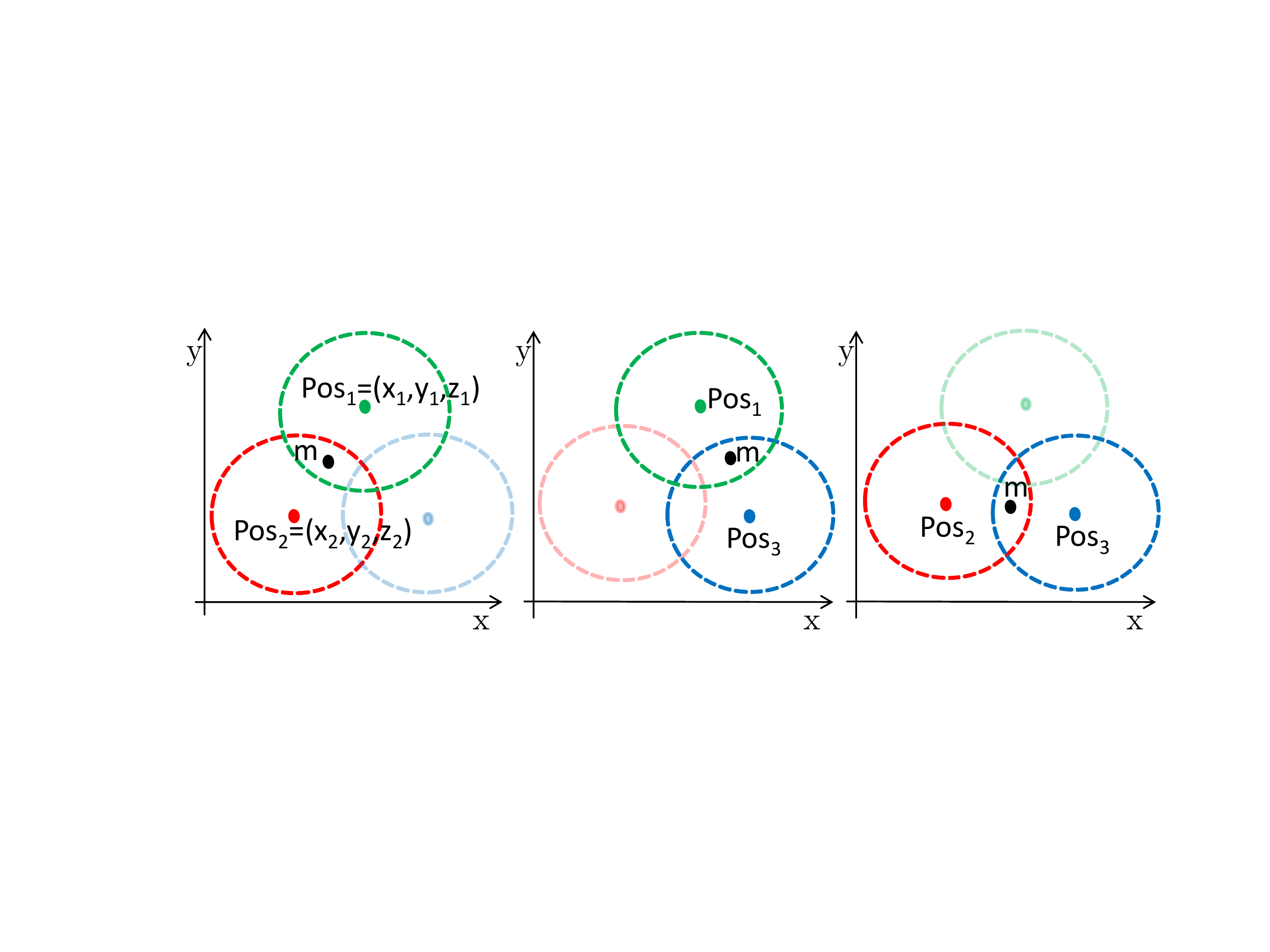}}
\caption{Example of the clustering algorithm for groups of 2 users (N = 2). User 1 in green, user 2 in red and user 3 in blue}
\label{fig_Clustering_Example_2}
\end{figure}

The algorithm to compute the common clusters is designed by AAU and it is based on finding the mean distance of the groups of users to the centroid of the groups. Then using a linear relationship (or another relationship) this distance gives a proportion of clusters to be shared among the group of users. This procedure is repeated for groups of two, three, four, etc. users until the maximum is reached. This algorithm is presented in Algorithm~\ref{alg_Algorithm} and Fig.~\ref{fig_Clustering_Example_2} shows an example.

\begin{algorithm}
 \ForEach{group of $N$ users ($N \in \{1,...,max(users)\}$)}{
  \eIf{$N = 1$}{
   proportion of clusters for the individual users $= 1$\;
   }{
   find centroid of the group of users: $m$\;
	\eIf{(all distances to $m$)$<$$R$}{
		find mean distance to $m$: $md$\;
		proportion of clusters $= \frac{-md}{R}+1$: $p$\;
		subtract $\frac{p}{N-1}$ from the groups containing $N-1$ users\;
		}{
			the users with (distance to $m$)$>$$R$ are too far away and no clusters are shared in this group\;
		}
  }
 }
 \caption{Compute the number of common clusters}
	\label{alg_Algorithm}
\end{algorithm}

In the previous algorithm the centroid of the group of users (i.e. $m$) is computed as,

\begin{equation}
m = \frac{(x_1,y_1,z_1)+\cdots+(x_N,y_N,z_N)}{N}
\end{equation}

where $(x_n,y_n,z_n)$ is the position of user $n$ in Cartesian coordinates (lets call it $Pos_n$).

Compute the distances from the users to the centroid and find if the auras are overlapping using:

\begin{equation}
\|m-Pos_n\|< R
\end{equation}

To compute the mean distance of the group of users to the centroid use:

\begin{equation}
md = \frac{\|m-Pos_1\|+\cdots+\|m-Pos_N\|}{N}
\end{equation}

The proportion of clusters to share corresponds to a linear relationship with the mean distance to the centroid (proportion of clusters $p=\frac{-md}{R}+1$). This linear function has been chosen for its simplicity. However, empirically derived cluster sharing functions can easily be substituted here.

\begin{figure}[!t]
\centering
\fbox{\includegraphics[trim = 2.4in 1.7in 2.2in 1.7in, clip, width=3.3in]{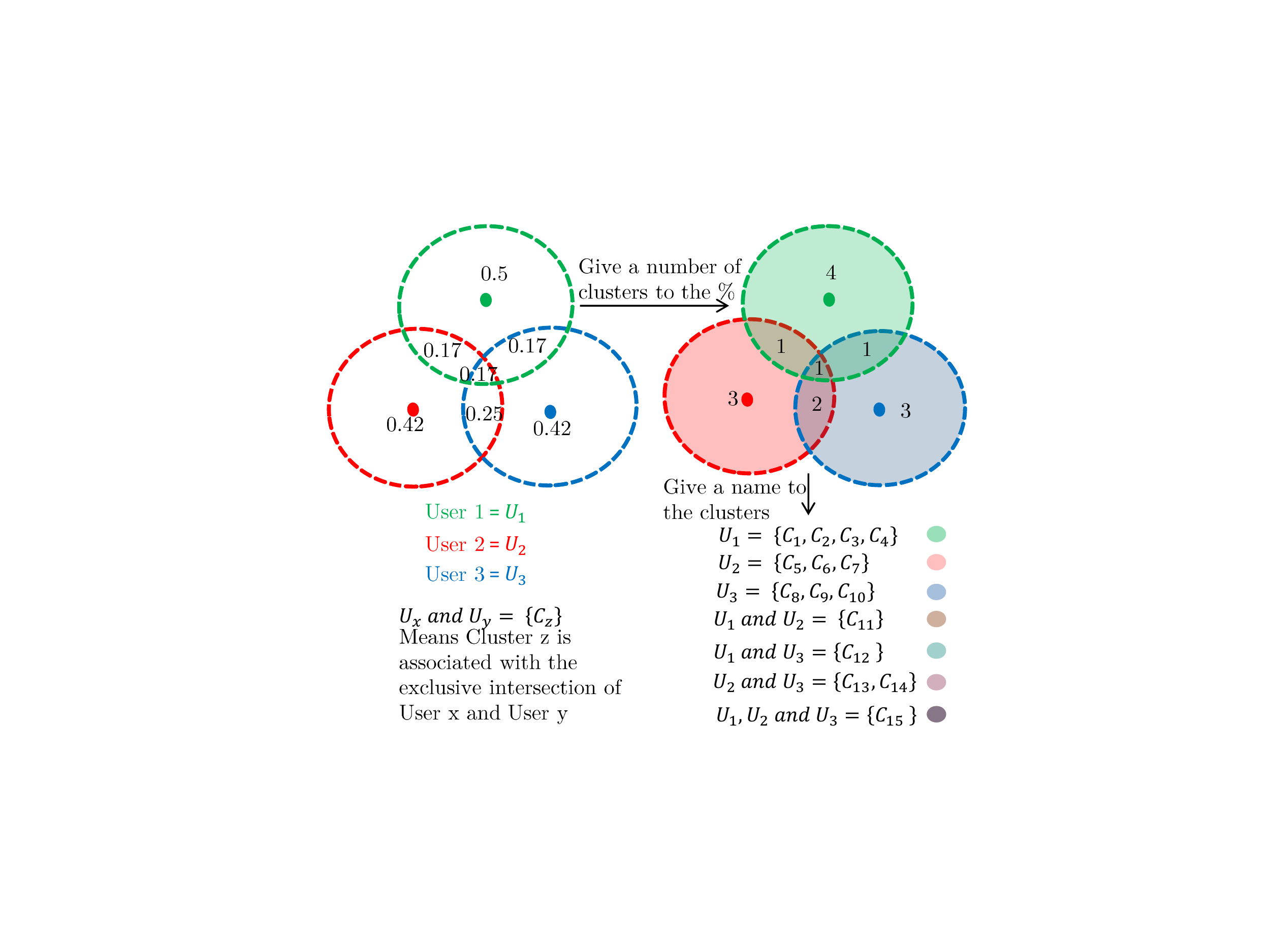}}
\caption{Example of the clustering algorithm with $7$ clusters (C) per user(U). First compute the proportion of common clusters for each group, then compute the number of clusters and finally assign a set of cluster names}
\label{fig_Clustering_Example}
\end{figure}

After the explained algorithm, each user has a proportion of individual clusters and each intersection of auras has a proportion of common clusters. Knowing the proportion of clusters to share and the total number of clusters, each user and group of users is assigned with a number of clusters as seen in Fig.~\ref{fig_Clustering_Example}. Notice that the parameters defining each cluster (i.e. angles, delay, position) have not been computed yet. The clusters are only defined by its number, and the parameters are computed in the following sub-step.

\subsection{Generation of initial parameters}
\label{sec_Generation}
The initial delays, powers and angles (i.e. azimuth of departure and arrival, elevation of departure and arrival) are generated for each cluster ($Cx$) in each segment following QuaDRiGa's procedure explained in~\cite{Jaeckel2014}. To create non-stationarities across the array, we modify this procedure to have one azimuth angle and one elevation angle of departure for each sub-array. There are $4 + 2A$ parameters for each cluster (being $A$ the number of sub-arrays). This procedure uses user ($Ux$) specific parameters (drawn from the large scale parameter maps) to generate the cluster parameters. If the cluster belongs only to one user (e.g. $C_3$ in Fig.~\ref{fig_Clustering_Example}) the parameters of that user are used to generate the cluster. On the other hand if the cluster belongs to more than one user (e.g. $C_{11}$ in Fig.~\ref{fig_Clustering_Example}) one of the users is picked to use its parameters to generate the cluster. We propose to pick the users randomly with a uniform distribution, but other methods are possible. The values of the departure angles are drawn independently for each sub-array.

\subsection{Computation of the focal points of user side clusters}
QuaDRiGa's drifting procedure determines the position of the Last Bounce Scatterer (LBS) and keeps it fixed during the whole segment,~\cite{Jaeckel2014}, global step 4. In this sub-step only the first part of the QuaDRiGa's drifting procedure is used to find the focal point of the LBS and add it to the table of parameters for each cluster. Even if the cluster belongs to more than one user (e.g. $C_{11}$ in Fig.~\ref{fig_Clustering_Example}) the focal point is referenced to the user used to generate the parameters of the cluster.

\begin{figure}[!t]
\centering
\fbox{\includegraphics[trim = 2in 2.1in 1.6in 2.2in, clip, width=3.3in]{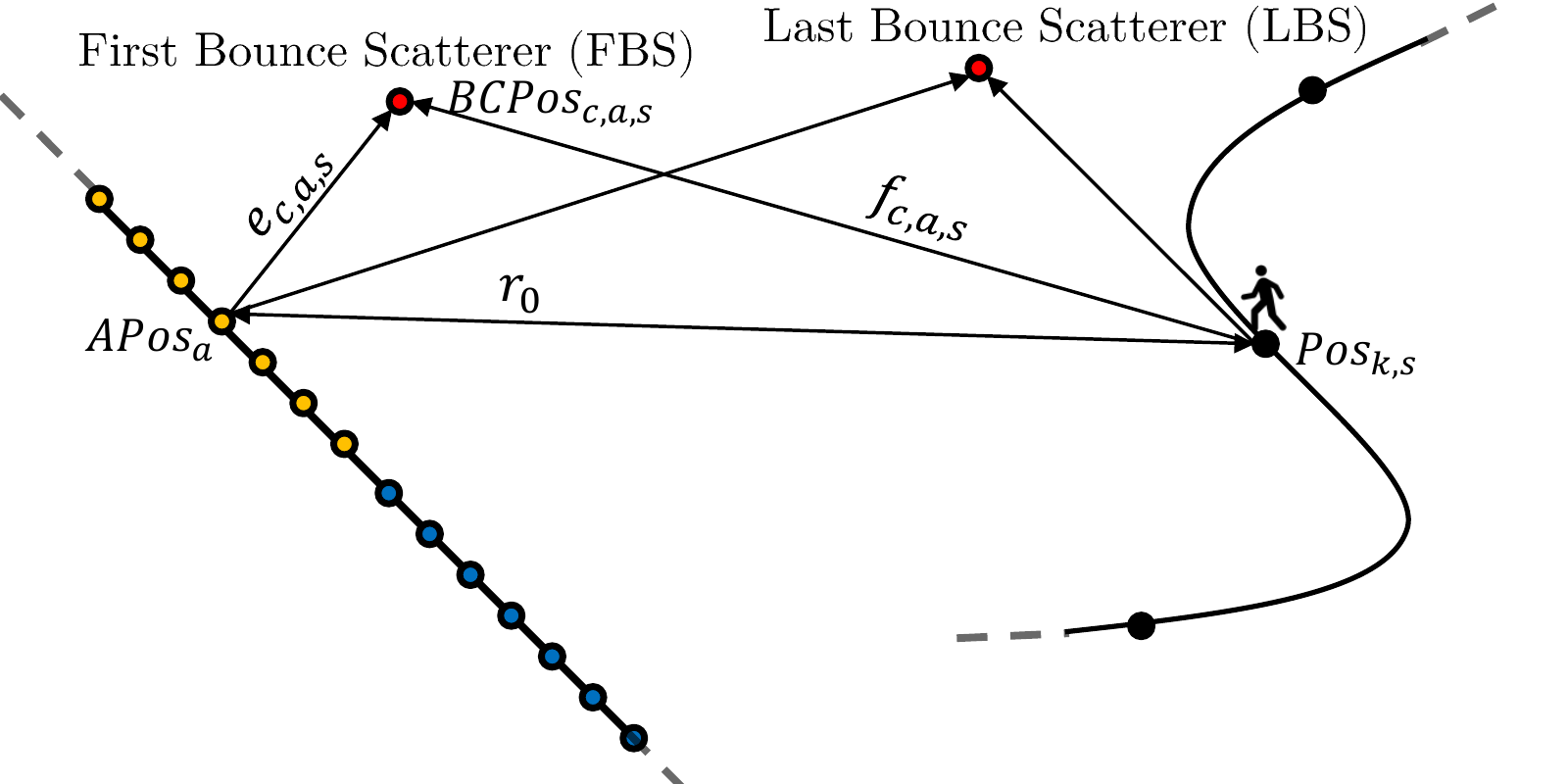}}
\caption{Computation of the focal point of the cluster at the transmitter side}
\label{fig_Cluster_Transmitter}
\end{figure}

Then it is necessary to find the focal point at the transmitter side. We call the focal point at the transmit side First Bounce Scatterer (FBS) analogous to the QuaDRiGa nomenclature. We propose to use the same procedure used by QuaDRiGa to find the focal point of the clusters at the transmit side. Next we explain how to adapt their procedure to the transmit side. Fig.~\ref{fig_Cluster_Transmitter} shows the parameters used.

First the total length (from transmitter, to cluster, to receiver) is obtained from the delay,

\begin{equation}
d_{c} = \tau_{c}c_0+|r_{0,a,k}|
\end{equation}
where $|r_{0,a,k}|$ is the distance between sub-array $a$ and user $k$ (i.e. $ ||APos_{a}-Pos_{k,s}||$, where $APos_{a}$ is the central position of the sub-array $a$ and $Pos_{k,s}$ is the first position of the user $k$ in the segment $s$), $\tau_{c}$ is the excess delay and $c_0$ is the speed of light.
Then the departure angles of the cluster are converted into Cartesian coordinates. $\hat{e}_{c,a,s}$ is the vector defining the direction of the cluster $c$ of sub-array $a$ at segment $s$.

$f_{c,a,s}$ defines the vector from the user to the cluster. Considering the triangle with vertices at the center of the sub-array, at the user position, and at the focal point of the cluster, and using the cosine theorem we can compute the distance from the sub-array to the cluster.

\begin{equation}
f_{c,a,s}^2 = |r_{0,a,k}|^2 + |e_{c,a,s}|^2 - 2|r_{0,a,k}||e_{c,a,s}|cos(\beta_{c,a,s})
\end{equation}

\begin{equation}
(d_c-|e_{c,a,s}|)^2 = |r_{0,a,k}|^2 + |e_{c,a,s}|^2 + 2|e_{c,a,s}|r_{0,a,k}^T \hat{e}_{c,a,s}
\end{equation}

\begin{equation}
|e_{c,a,s}| = \frac{d_c^2-|r_{0,a,k}|^2}{2(d_c r_{0,a,k}^T \hat{e}_{c,a,s})}
\end{equation}

The vector from the transmitter position to the focal point of the cluster at the transmit side is,

\begin{equation}
e_{c,a,s} = |e_{c,a,s}|\hat{e}_{c,a,s}
\end{equation}

And using this vector and the position of the transmit array we can find the position of the focal point of the cluster at the transmit side (FBS) as
\begin{equation}
BCPos_{c,a,s} = e_{c,a,s} + APos_a
\end{equation}

After adding the focal points, the clusters have 5+3A parameters in their tables (i.e. power, delay, azimuth of arrival, elevation of arrival, focal point at receiver and for each sub-array: azimuth of departure, elevation of departure and focal point at transmitter).

\subsection{Sharing the clusters}
\label{sec_Sharing}
\begin{figure}[!t]
\centering
\includegraphics[trim = 1.8in 3.1in 2.4in 2.7in, clip, width=3.3in]{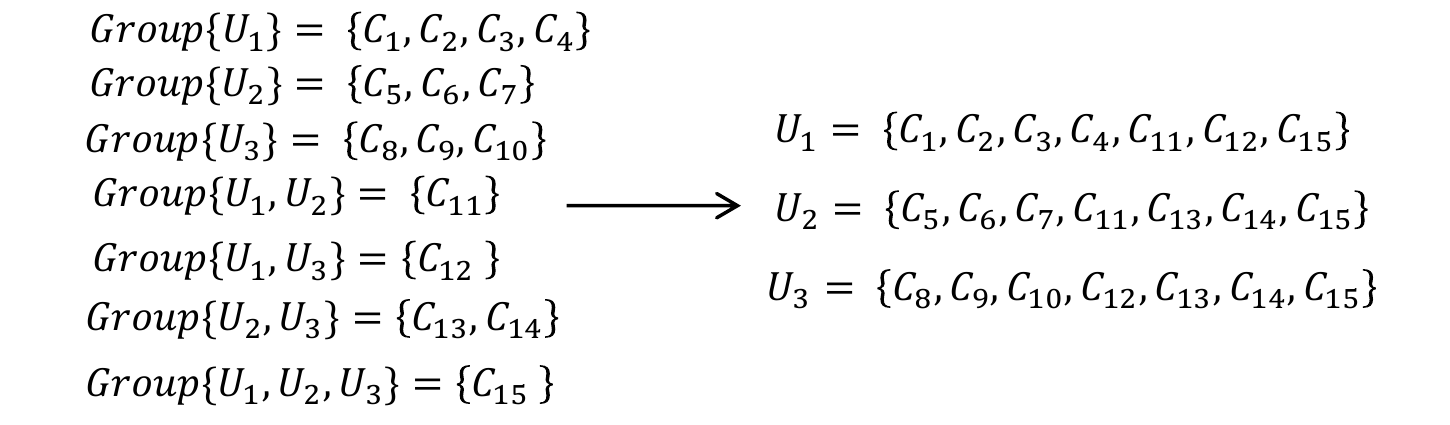}
\caption{Example of the cluster sharing}
\label{fig_Cluster_sharing}
\end{figure}

In sub-step \ref{sec_Calculate} we showed that each cluster could have more than one owner, but in sub-step \ref{sec_Generation} the parameters of only one user have been used to generate the cluster. In this sub-step the clusters are shared with the other users that they belong to, according to the results of~\ref{sec_Calculate}. In other words, the clusters and their generated parameters are duplicated to the parameter tables of the corresponding users. See an example in Fig.~\ref{fig_Cluster_sharing}.

\subsection{Recalculating parameters}
\begin{figure}[!t]
\centering
\includegraphics[trim = 1.1in 2.6in 1.1in 2.6in, clip, width=3.3in]{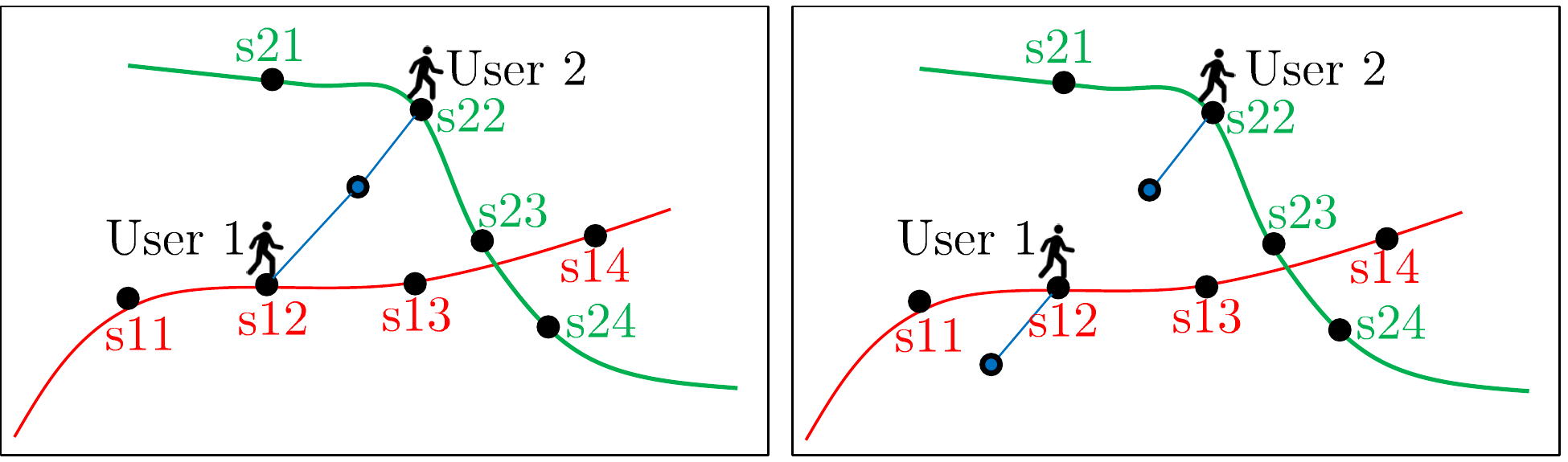}
\caption{If the cluster is near to the users to share the parameters can result in very different clusters (right). It is better to share the focal point (left)}
\label{fig_Recalculating}
\end{figure}

As some clusters have been generated using the parameters of one user, but after sub-step \ref{sec_Sharing} they have been shared with another user, it is necessary to recalculate the parameters for the new user. The reason is because the focal point of the cluster has been calculated using the angles and positions of one user, but the position of the other user can be different. There are two options to recalculate the parameters, both shown in Fig.~\ref{fig_Recalculating}. The first option is to keep the same parameters generated in sub-step $3$ and recalculate the two focal points of the cluster for the new user. The second option is to keep the same focal point and recalculate the other parameters (including the angles of departure). If the clusters are far away from the users it is possible to keep the same parameters and avoid recalculating the focal point because the relative position does not change very much. However, if the cluster is near the users, we have to recalculate the focal point, else it would result in effectively different clusters for the users. We propose that if the clusters are less than $3$ segment lengths away the focal point is kept and the other parameters are recalculated. Otherwise, the opposite happens.

\section{Conclusion}
This paper extends the framework of the existing Winner-type GSCM towards the evolution of 5G channel models for massive MIMO. Winner-type GSCM are heavily employed by the industry, so a modification is necessary to continue build on existing knowledge base. The paper focuses on the three main limitations of the existing models that prevents the proper simulation of massive MIMO systems. First of all the lack of a method to model the multiuser consistency. Then, the impossibility to generate non-stationarities over the base station array. Finally the limitation of using the planar wave approximation. Using QuaDRiGa as a reference model, several modifications are proposed to overcome these limitations.

\section*{Acknowledgment}
The research presented in this paper was partly supported by the
Danish Council for Independent Research (Det Frie Forskningsr{\aa}d)
DFF–1335–00273. This work was supported by Huawei Technologies Co. Ltd. (Huawei Sweden) in the framework of the cooperation project No. 861020.

\bibliographystyle{IEEEtran}
\bibliography{BIB_PIMRC}

\end{document}